\documentclass{mn2e}
\usepackage{epsfig}
\usepackage{times}
\begin{document}

\title[] {
Towards a unified model for black hole X-ray binary jets
}
\author[Fender, Belloni \& Gallo]  {R. P. Fender$^1$, T. M. Belloni$^2$, E. Gallo$^1$ \\
$^1$Astronomical Institute
`Anton Pannekoek', University of Amsterdam, and Center for High Energy
Astrophysics, Kruislaan 403, \\ 1098 SJ, Amsterdam, The Netherlands
{\bf rpf@science.uva.nl, egallo@science.uva.nl}\\ 
$^2$ INAF -- Osservatorio Astronomico di Brera, Via E. Bianchi 46, I-23807
Merate, Italy {\bf belloni@merate.mi.astro.it}
}

\maketitle

\begin{abstract}
We present a unified semi-quantitative model for the disc-jet coupling
in black hole X-ray binary systems. In the process we have compiled
observational aspects from the existing literature, as well as
performing new analyses. We argue that during the rising phase of a
black hole transient outburst the steady jet known to be associated
with the canonical 'low/hard' state persists while the X-ray spectrum
initially softens. Subsequently, the jet becomes unstable and an
optically thin radio outburst is always associated with the soft X-ray
peak at the end of this phase of softening. This peak corresponds to a
'soft very high state' or 'steep power law' state. Softer X-ray states
are not associated with 'core' radio emission. We further demonstrate
quantitatively that the transient jets associated with these optically
thin events are considerably more relativistic than those in the
'low/hard' X-ray state. This in turn implies that as the disc makes
its collapse inwards the jet Lorentz factor rapidly increases,
resulting in an internal shock in the outflow, which is the cause of
the observed optically thin radio emission. We provide simple
estimates for the efficiency of such a shock in the collision of a
fast jet with a previously generated outflow which is only mildly
relativistic. In addition, we estimate the jet power for a number of
such transient events as a function of X-ray luminosity, and find them
to be comparable to an extrapolation of the functions estimated for
the 'low/hard' state jets. The normalisation may be larger, however,
which may suggest a contribution from some other power source such as
black hole spin, for the transient jets. Finally, we attempt to fit
these results together into a coherent semi-quantitative model for the
disc-jet coupling in all black hole X-ray binary systems.
\end{abstract}


\section{Introduction}

Relativistic jets are a fundamental aspect of accretion onto black
holes on all scales. They can carry away a
large fraction of the available accretion power in collimated flows
which later energise particles in the ambient medium. The removal of
this accretion power and angular momentum must have a dramatic effect
on the overall process of accretion. In their most spectacular form
they are associated with supermassive black holes in active galactic
nuclei (AGN), and with Gamma-Ray Bursts (GRBs), the most powerful and
explosive engines in the Universe respectively. However, parallel
processes, observable on humanly-accessible timescales, are occurring
in the accretion onto black holes and neutron stars in binary systems
within our own galaxy.

The current observational picture of X-ray binary jets is most simply put as
follows: in the low/hard state, which exists typically below a few \%
of the Eddington luminosity $L_{\rm Edd}$ (e.g. Maccarone 2003;
McClintock \& Remillard 2004) there is a `compact' self-absorbed jet
which manifests itself as a `flat' (spectral index $\alpha \sim 0$
where $\alpha = \Delta \log{S_{\nu}} / \Delta \log{\nu}$) or
`inverted' ($\alpha \geq 0$) spectral component in the radio,
millimetre and (probably) infrared bands (e.g. Fender 2001b, Corbel \&
Fender 2002). The radio luminosity of these jets shows a strong,
non-linear correlation with X-ray luminosity (Corbel et al. 2003;
Gallo, Fender \& Pooley 2003) and has only been directly spatially
resolved in the case of Cyg X-1 (Stirling et al. 2001), although the
`plateau' jet of GRS 1915+105 is phenomenologically similar and has
also been resolved (Dhawan et al. 2000; Fuchs et al. 2003). The
suggestion that such steady, compact jets are produced even at very
low accretion rates (Gallo, Fender \& Pooley 2003; Fender, Gallo \&
Jonker 2003) has recently received support in the flat radio spectrun
observed from the `quiescent' transient V404 Cyg at an average X-ray
luminosity $L_{\rm X} \sim 10^{-6} L_{\rm Edd}$ (Gallo, Fender \&
Hynes 2004). During steady 'soft' X-ray states the radio emission, and
probably therefore jet production, is strongly suppressed (Tanabaum et
al. 1972; Fender et al. 1999b; Corbel et al. 2001; Gallo, Fender \& Pooley 2003).

Additionally there are bright events associated with transient
outbursts and state transitions (of which more later), which are often
directly resolved into components displaying relativistic motions away
from the binary core (e.g. Mirabel \& Rodriguez 1994; Hjellming \&
Rupen 1995; Fender et al. 1999) not only in the radio but also -- at
least once -- in the X-ray band (Corbel et al. 2002). These events
typically display optically thin (synchrotron) radio spectra ($\alpha
\leq -0.5$). Both kinds of jets are clearly very powerful and coupled
to the accretion process. See Mirabel \& Rodriguez (1999) and Fender
(2004) for a more thorough review of the observational properties of
X-ray binary jets.

In this paper we attempt to pin down as accurately as possible the
moment at which the major radio outburst occurred and relate this to
the X-ray state at the time. We subsequently compare this with the
X-ray state corresponding to the lower-luminosity steady jets, to the
evolution of transient outbursts, and to the velocity and power
associated with each 'type' of jet, in order to draw up a framework
for a unified model of black hole X-ray binary jet production.

Several black hole systems are investigated in this paper, and in
addition we compare these with the neutron star systems Cir X-1 and
Sco X-1. The data relating to the radio flares, jet Lorentz factors
(if measured), corresponding X-ray luminosities, estimated distances
and masses, are summarised in Table 1.

\begin{table*}
\begin{tabular}{cccccccccc}
\hline
Source & $d$  &$M$ & Date & $\Delta t$ & $S_{\rm 5GHz}$ & $\Gamma$& $L_{\rm J}$ &$L_{\rm X, VHS}$ & Ref \\
       & (kpc) & $M_{\odot}$ & (MJD) & (sec) & (mJy) & & (Edd) & (Edd) &\\
\hline
GRS 1915+105 (flare) & 11 & 14 & 50750 & 43200& 320  & $\leq 1.4$ &0.6& 1.1 &F99 \\
GRS 1915+105 (osc.) & 11 & 14 & many & 300& 50  & $\geq 2$ &0.05& 1.1& F99 \\
GROJ1655-40 & 3.5 & 7 & 49580 & 43200 & 2000 & $\geq 1.7$ & 1.0 &0.1 &HR95\\
XTE J1859+226 & 6 & 7 & 51467 & 21600 & 50 & ? &0.2&0.2 &B02 \\
XTE J1550-564 & 6 & 9 & 51077 & 43200 & 130 & $\geq 2$ &0.3& 0.5&W02 \\
GX 339-4 & 8 & 7 & 52408 & 19800 & 55 & $\geq 2.3$ &0.3&0.07 &G04 \\
V4641 Sgr & 8 & 9 & 51437 & 43200 & 420 & $\geq 9.5$ &0.8& 4&Hj01, Or01\\
Cyg X-1 & 2.5 & 10 & 53055 & 2000 & 50 & ? & 0.02 & $\sim 0.05$ &P04 \\
XTEJ1748-288 & 8.5 & 7 & 50980 & 172800 & 530 & $\geq 2.7$ & 1.9& $\sim 0.1$ &B04 \\ 
\hline
Cir X-1 & 6 & 1.4 & 51837 & 43200 & 20 & $\geq 15$ & 0.6 & $\sim 0.7$ &F04 \\
Sco X-1 & 2 & 1.4 & many & 3600 & 15 & $\geq 3.2$ & 0.02 & $\sim 1$ &F02 \\
\hline
\end{tabular}
\caption{Parameters for the X-ray binary systems and selected jet
  events discussed in this paper; the last two sources contain
  accreting neutron stars, the rest black hole candidates. The first
  three columns give the source name, distance and mass of the
  accretor. Columns 4--9 give the date, rise time, peak radio flux,
  constraints on bulk Lorent factor, estimated jet power and
  corresponding estimated X-ray power for jet production events. The
  final column gives references (F99 = Fender et al. 1999; HR95 =
  Hjellming \& Rupen 1995; B02 = Brocksopp et al. 2002; W02 = Wu et
  al. 2002; G04 = Gallo et al. 2004; Hj01 = Hjellming et al. 2001;
  Or01 = Orosz et al. 2001; P04 = Pooley 2004; B04 = Brocksopp et
  al. (in prep); F04 = Fender et al. 2004; F02 = Fomalont et al. 2001a,b).
}
\end{table*}

\section{The sample: four black holes undergoing jet formation}

Is there a signature in the X-ray light curve of an outbursting source
which indicates when the relativistic jet is launched ? In the
following we investigate radio and X-ray light curves of four black
hole binaries -- GRS 1915+105, GX 339-4, XTE J1859+226 and XTE
J1550-564 -- undergoing state transitions in order to investigate this
point. 

\subsection{X-ray Data analysis}
For XTE J1550-564, XTE J1859+226 and GX 339-4, we extracted the
background-subtracted PCA count rate, using PCU2 only, for the each
available RXTE observation relative to the first part of the outburst
considered. For each observation, we also produced an X-ray color (or
hardness ratio) by dividing the background-subtracted count rates in
the 6.3-10.5 keV band by those in the 3.8-6.3 keV band.  For GRS
1915+105, given the much shorted time scales involved, we analyzed a
single observation (see Sect. 2.2.1), producing a PCA light curve at 1
second resolution from all PCUs summed together, and an X-ray color
curve at the same time resolution. The energy bands used for the
color, 15.2-42.3 keV and 2.1-5.9 keV were different. The reason for
which is that the thermal disc component in GRS 1915+105 is
considerably higher than in other systems (see Fender \& Belloni
2004): using the same energy bands results in the harder C state
having a softer color than the softer B state. The use of a harder
energy band for the numerator ensures that the thermal disc component
does not strongly contaminate it.

In order to estimate in a homogeneous way the source luminosity at
peak of the outburst for XTE J1550-564, XTE J1859+226 and GX 339-4, we
extracted RXTE PCA+HEXTE spectra from the public archive corresponding
to the peak flux in the PCA band. Spectra were created from PCU2 data
(3-25 keV) for the PCA and from cluster A data (20-200 keV) for HEXTE,
using FTOOLS 5.3. The spectra were corrected for background and
dead-time effects. A 0.6\% systematic error was added to the PCA to
account for residual calibration effects.  We fitted the spectra with
the standard phenomenological model for these systems, consisting of a
cutoff power-law ({\tt cutoff}), a disk blackbody ({\tt diskbb}), a
Gaussian emission line between 6 and 7 keV, all modified by
interstellar absorption. The actual models might be more complex, but
we are interested in the determination of the flux only, so that the
presence of additional components like iron absorption edges does not
change significantly our results. The reduced $\chi^2$ values were
between 1.3 and 1.5.  From the best fit models for each source, in
order to estimate the bolometric flux, we computed the unabsorbed flux
of the disc blackbody component in the 0.001-100 keV range and that
of the power-law component in the 2-100 keV range.  For GRS 1915+105,
given the much shorter time scales involved for the short
oscillations, we applied the procedure outlined above to the
RXTE/PCA+HEXTE data of the observation of 1997 October 25, the last
one before the launch of the major jet observed with MERLIN (Fender et
al. 1999).  In order to approximate the flux evolution from the count
rate light curves, we applied to all preceding and following
observations the count rate to flux conversion factor obtained from
the peak. A comparison with published flux values for XTE J1550-564
(Sobczak et al. 2000) this proved to be a reasonable approximation for
our purposes.

Much of the following discussion is based upon the association of
radio emission with the X-ray 'states' of different sources. We will
use the following abbreviations throughout the paper: the 'low/hard'
state as 'LS'; the 'high/soft' state as 'HS', and the 'very high' or
'intermediate' states as 'VHS/IS'. In fact, as we shall discuss
further (see also Belloni et al. 1994; Homan \& Belloni 2004), the
VHS/IS is not a single state but has both 'hard' (as used to describe
the X-ray spectrum) and 'soft' varieties.

Note that recently McClintock \& Remillard (2004) have proposed a
modification to this classification scheme in which the LS is referred
to as the 'hard' state and the HS as the 'thermal dominated state',
while revised definitions are introduced for the VHS/IS states.  While
we retain the 'classical' definitions of states, we discuss later on
how these correspond to the revised state definitions of McClintock \&
Remillard.

\subsection{Individual sources}

\begin{figure*}
\label{lc}
\centerline{\epsfig{file=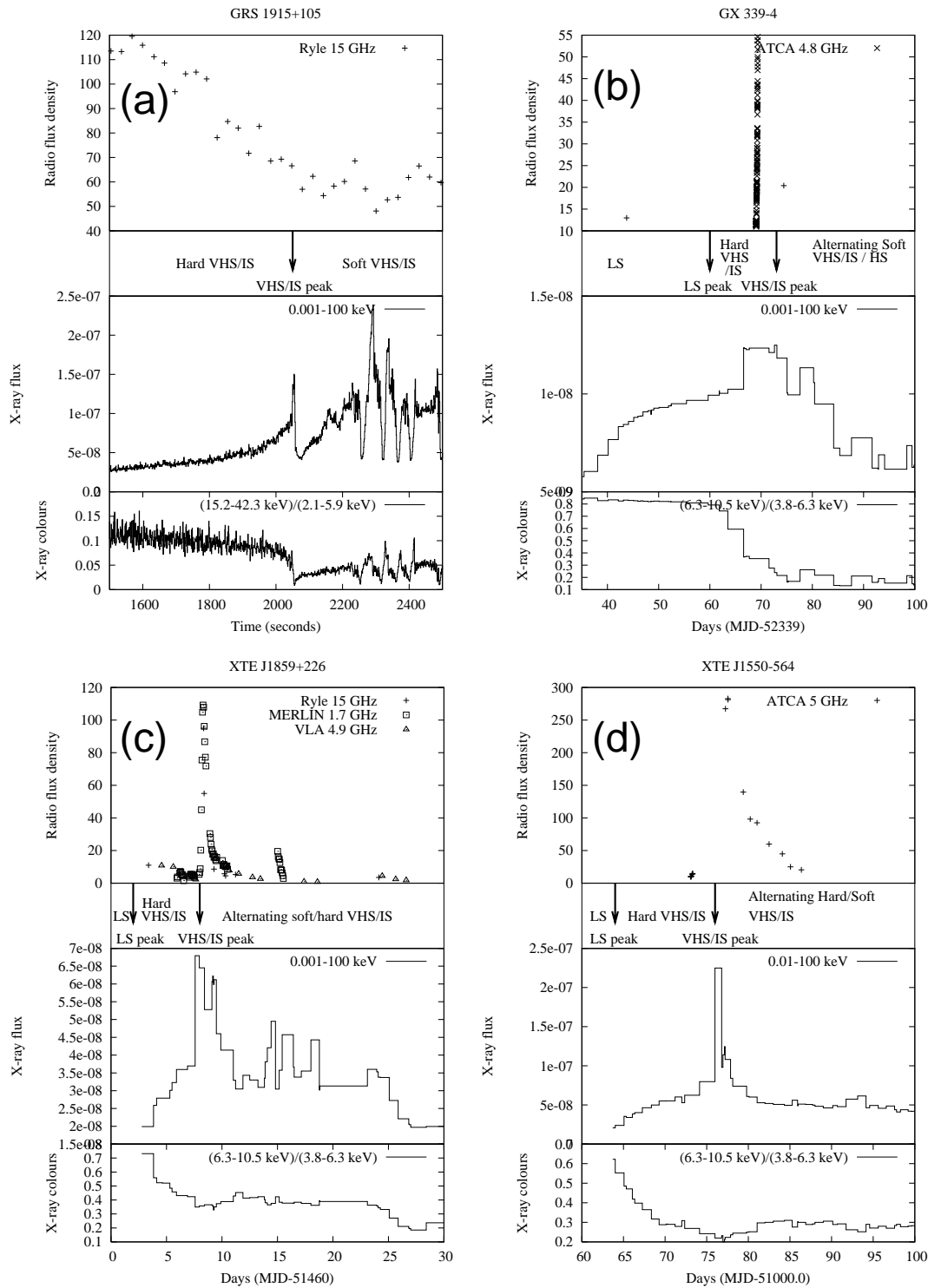, angle=0, width=17cm}}
\caption{Radio and X-ray light curves, X-ray colours and X-ray state
  classifications during periods around transient jet formation, for
  four black hole (candidate) X-ray binaries. In GRS 1915+105 the
  canonical LS or HS are never reached; in GX 339-4, XTE J1859+226 and XTE
  J1550-564 the delay between the canonical LS peak and subsequent
  VHS/IS peak ranges from a few days to two weeks. Nevertheless, in
  all four cases the radio flare occurs at the time of the VHS peak,
  indicating a clear association between this, and not the previous
  LS, and the major ejection. The units of the X-ray flux are erg
  s$^{-1}$ cm$^{-2}$.}
\end{figure*}

\begin{figure*}
\label{1859xrad}
\centerline{\epsfig{file=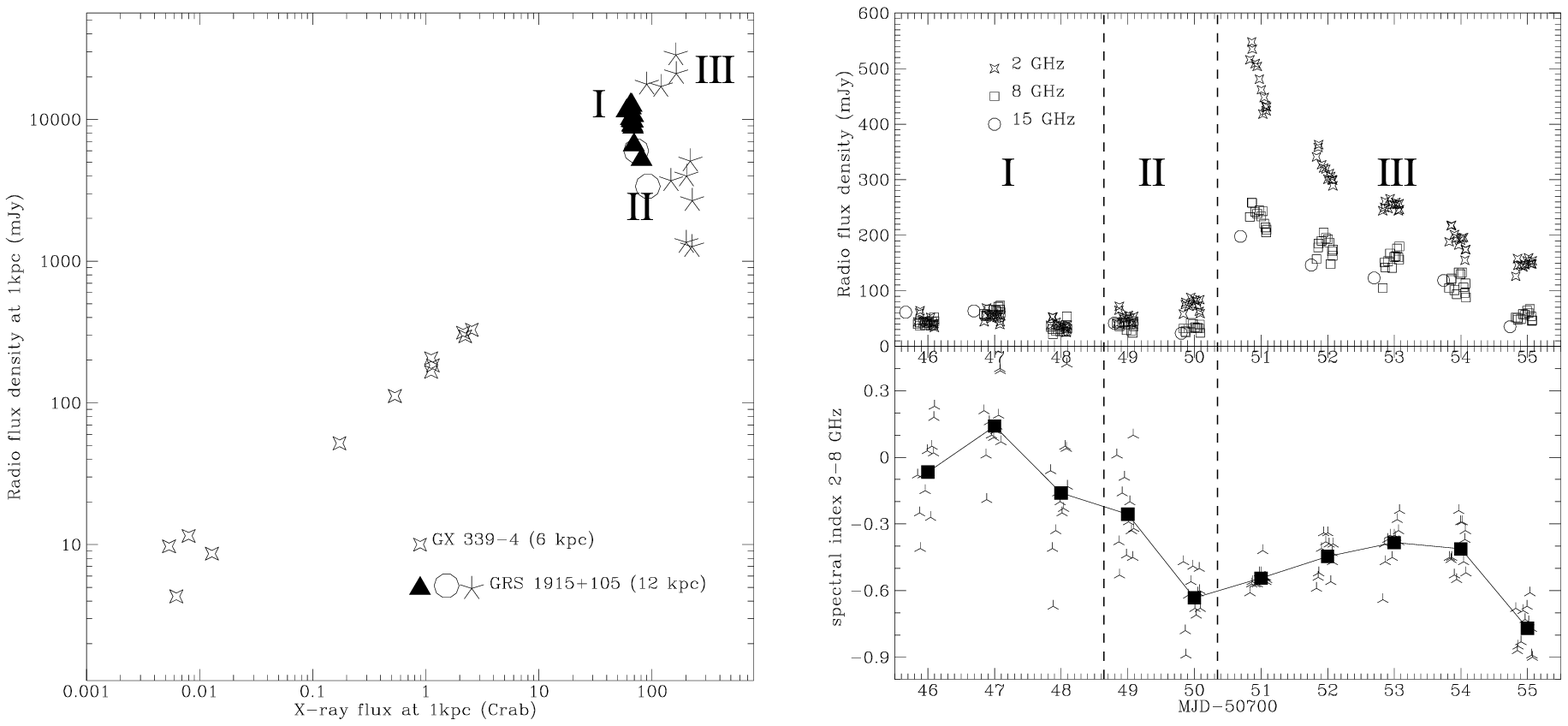, angle=0, width=16cm}}
\centerline{\epsfig{file=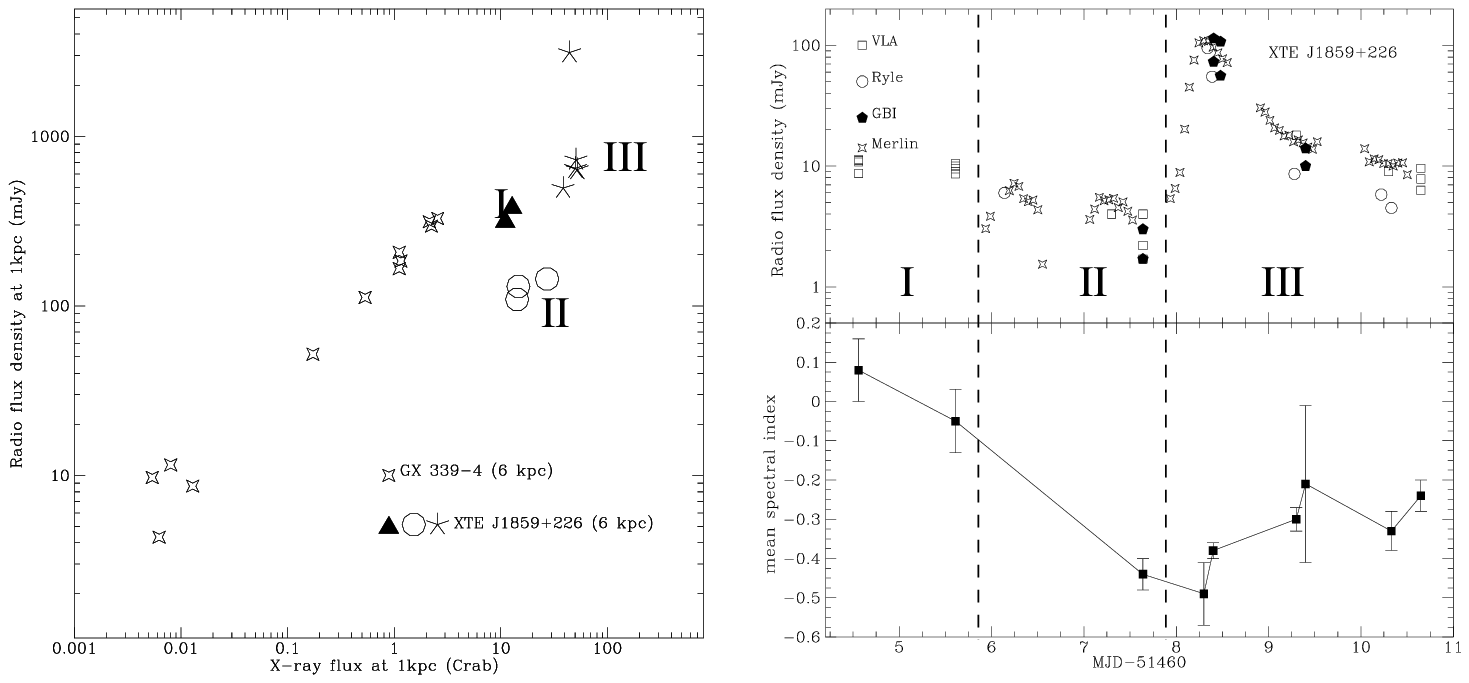, angle= 0, width=16cm}}
\caption{Behaviour of radio emission immediately prior to optically
  thin flare events. The top panel shows data for the GRS 1915+105
  plateau reported in detail in Fender et al. (1999); the lower panel
  the outburst of XTE J1859+226 reported in Brocksopp et al. (2002)
  and also our Fig 1.  The right-hand panels show in detail the radio
  light curves and fitted spectral indices over this period, and are
  separated for each source into three phases. The location of these
  phases in the radio luminosity : X-ray luminosity ($L_{\rm
  radio}$:$L_{\rm X}$) plane (Gallo, Fender \& Pooley 2003) is
  indicated in the left-hand panels (the X-ray luminosity is
  calculated from the RXTE ASM, as in Gallo, Fender \& Pooley 2003,
  and is therefore much less accurate than the fluxes presented in Fig
  1).  Initially, while still in a 'hard' VHS/IS, both sources display
  optically thick radio emission which lies close in the ($L_{\rm
  radio}$:$L_{\rm X}$) plane to the mean relation for LS BH XRBs, as
  marked out by the data for GX 339-4 (Gallo, Fender \& Pooley
  2003). Subsequently, the radio emission seems to become more erratic
  and occasionally optically thin.  }\end{figure*}

\subsubsection{GRS 1915+105}

GRS 1915+105 has long been a key source in our understanding of the
disc-jet coupling in X-ray binary systems (see Fender \& Belloni 2004
and references therein). In the context of this study it is
interesting because its X-ray state never seems to reach the
'canonical' LS or HS but instead switches between 'hard' (state 'C' of
Belloni et al. 2000) and 'soft' (states 'A' and 'B') VHS/IS states
(see also Reig et al. 2003).  It has clearly been established in this
source that phases of hard (C) X-ray emission lasting more than a few
hundreds of seconds are associated with radio events, and that when
the source is only in soft (A,B) states there is no radio emission
(Klein-Wolt et al. 2002). In the context of this work its importance
is therefore in establishing that state changes 'within' the VHS/IS
can produce radio outbursts without requiring any 'contact' with the
canonical LS or HS. Fig 1(a) presents the lightcurve of a typical
'oscillation' event in which the source makes a transition from state
C to state A/B.  The data correspond to the RXTE observation of 1999
June 14 (class $\nu$), time zero is 01:00:40 UT.These oscillation
events can occur in very long sequences and are generally associated
with sequences of synchrotron oscillations which are almost certainly
from the jets (e.g. Klein-Wolt et al. 2002).

\subsubsection{GX 339-4}

GX 339-4 is also a key source in our understanding of the disc-jet
coupling in X-ray binaries (Fender et al. 1999b; Corbel et al. 2000,
2003; Gallo et al. 2003; Belloni et al. 2004), albeit one which varies
on timescales considerably longer than in GRS 1915+105. Recently, a
clear bright optically thin radio flare was observed from this source,
and subsequently found to be associated with a relativistic ejection
event (Gallo et al. 2004). The light curve, corresponding to the first
part of the 2002/2003 outburst (Belloni et al. 2004), presented in Fig
1(b) shows many similarities with that of GRS 1915+105 (Fig 1(a)) in
the X-ray band, with a rising hard (in this case the canonical LS)
state softening shortly before a soft VHS/IS peak.  Subsequently both
sources show a drop in the X-ray flux and then a slow recovery to even
higher levels. Compared to GRS 1915+105 we see that the softening of
the X-ray state began a few days {\em before} the radio flare, which
seems to correspond to the VHS/IS peak near the {\em end} of the
softening.  However, the key thing we learn from GX 339-4 compared to
GRS 1915+105 is that the major radio event is also associated with the
state transition, something which is not unambiguous in GRS
1915+105. In this context it is interesting to note that the rise and
decay timescales of the radio oscillation events in GRS 1915+105 are
comparable to the durations of both the hard VHS/IS and
soft VHS/IS states (since the source is in general 'oscillating'
between the two). However, in GX 339-4 we see that this is clearly not
the case, and that the transient radio event is associated with a
specific and very limited instant in time.

\subsubsection{XTE J1859+226}

XTE J1859+226 (Fig 1(c)) is a more 'traditional' transient than either
GRS 1915+105 or GX 339-4, and underwent a bright outburst in 1999
(Wood et al. 1999; Markwardt et al. 1999; Brocksopp et al. 2002;
Casella et al. 2004).
It showed an initial LS peak during the
rise to outburst, which preceded the subsequent soft peak by several
days. This allows us to separate the LS and VHS/IS peaks -- which was
impossible for GRS 1915+105 and still difficult for GX 339-4 -- and
identify which is associated with the radio event. The radio data used
in Fig 1(c) are from Brocksopp et al. (2002).

This source peaked in hard X-rays (see BATSE light curve in Brocksopp
et al. 2002) about a week before the soft peak. The X-ray state
between the two peaks can be formally described as a 'hard VHS/IS'
which is
gradually softening. The optically thin radio event is clearly
associated not with the peak of the LS, after which the X-ray
spectrum starts to soften, but with the VHS/IS peak, which seems to
occur just at the end of the spectral transition. Note also that some
radio emission persists after the LS peak, into the VHS/IS (further
discussion below). 

\subsubsection{XTE J1550-564}

XTE J1550-564 (Fig 1(d)) is another bright transient which has
undergone several outbursts in recent years. The data analysed here
correspond to the brightest one, in 1998/1999 (see Sobczak et
al. 2000; Homan et al. 2001; Remillard et al. 2002).  This outburst
was associated with a very strong optically thin radio event
subsequently resolved into a radio (Hannikainen et al. 2001) and, most
spectacularly, an X-ray jet moving relativistically (Corbel et
al. 2002). The radio data plotted in Fig 1(d) are from Wu et
al. (2002) and Hannikainen et al. (2004).

As in XTE J1859+226, the LS and VHS/IS peaks are clearly separated in
time, in this case by approximately two weeks. Cursory inspection of
Fig 1(d) clearly indicates that the optically thin jet is launched at
the time of the VHS/IS peak, which occurs, again, just at the end of
the X-ray spectral transition. Once more, as in XTE J1859+226, we also
note low-level radio emission between the LS and VHS/IS peaks.
 
\section{Jets as a function of X-ray state: new perspectives}

Based upon the investigation we have performed, we are better able to
associate the characteristics of the radio emission as a function of
X-ray state, and therefore to probe the details of the jet:disc
coupling. While the previously-established pattern of:

\begin{itemize}
\item{LS = steady jet}
\item{HS = no jet}
\end{itemize}

remains valid, additional information has clearly come to light about
the details of jet formation in the VHS/IS during transient outbursts.

\subsection{Persistence of the steady jet in the 'hard' VHS/IS}

It has by now been established for several years that the canonical LS
is associated with persistent flat- or inverted-spectrum radio
emission which probably arises in a self-absorbed, self-similar
outflow (Fender 2001). This radio emission is correlated with the
X-ray emission as $L_{\rm radio} \propto L_{\rm X}^{b}$ where $b \sim
0.7$ with a small apparent range in normalisations for several
different systems (Corbel et al. 2003; Gallo, Fender \& Pooley 2003;
see also Fender, Gallo \& Jonker 2003 and Jonker et al. 2004 for
further discussions and implications). 

While the HS was known to be associated with a dramatic decrease in
the radio emission (e.g. Tananbaum et al. 1972; Fender et al. 1999b;
Gallo, Fender \& Pooley 2003) it was not well known how the radio
emission behaved in the VHS/IS. Fender (2001a) suggested, based upon
GRS 1915+105, that the VHS/IS was associated with unstable, discrete
ejection events. Corbel et al. (2001) reported that the radio emission
from XTE J1550-564, eleven days after a transition to the VHS/IS, was
reduced ('quenched') by a factor of 50 compared to the previous LS. 

However, it is clear from Fig 1 that some radio emission persists
beyond the end of the canonical LS, before the outburst. Some of the
best data available are those of XTE J1859+226 (Brocksopp et
al. 2002), which are plotted in Fig 1(c) and Fig 2. For at least three
days following the LS peak, while the X-ray spectrum is softening, the
radio emission is persistent with an approximately flat
spectrum. Furthermore, the radio luminosity remains consistent with
the universal relation of Gallo, Fender \& Pooley (2003); this is
demonstrated in the left panels of Fig 2.  In fact, inspecting Fig 3
of Gallo, Fender \& Pooley (2003) we can see that in the case of Cyg
X-1 the turnover in the radio emission occurs {\em after} there has
already been some softening of the X-ray emission, implying also that
the compact jet persists into the 'hard VHS/IS'. Based on a recent
study of XTE J1650-500, Corbel et al. (2004) have also concluded that
the steady jet may persist beyond the initial softening of the
canonical LS. The importance of this is that there is clearly {\em
not} a one-to-one relation between the behaviour of the radio emission
and the X-ray states as currently defined (this will be discussed more
later).

\subsection{Changes in the jet radio spectrum {\em prior} to the outburst}

Following the persistence of the steady LS-like radio emission into
the hard VHS/IS, the data indicate that a change in the radio emission
does occur prior to the radio flare. In brief, it appears that the
radio emission starts to become more variable, with a peaked or (more)
optically thin spectrum shortly before the radio flare. We present the
evidence for this from the four key sources under study here:

\begin{figure}
\label{allthree}
\centerline{\epsfig{file=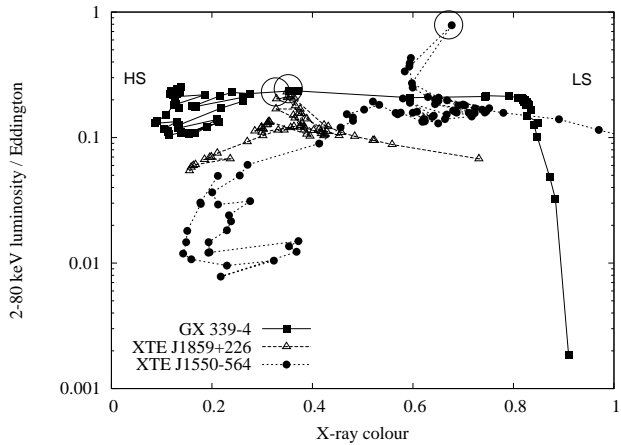, angle=270, width=9.5cm}}
\caption{Combined X-ray hardness-intensity diagram (HID) for GX 339-4,
  XTE J1859+226 and XTE J1550-564. The X-ray fluxes plotted in Fig
  1(b--d) have been scaled to Eddington-ratioed luminosities using the
  distance and mass estimates given in table 1. Note that ejections in
  GX 339-4 and XTE J1859+226 occur at almost exactly the same colour
  and X-ray luminosity. Most of the data points correspond to varying
  degrees of the VHS/IS, and not the canonical LS (to the right) or HS
  (to the left).  }
\end{figure}

\subsubsection{GRS 1915+105}

Given that in GRS 1915+105 we (i) are typically observing radio
emission from the {\em last} state transition, (ii) the transition
period between the 'hard' and 'soft' VHS/IS is very rapid, it is not
possible to investigate this effect for the oscillation-type events in
this source. However, the effect is seen in the monitoring of the
plateau state and subsequent radio flares as reported in Fender et
al. (1999a). As noted in that work (see also discussion in Klein-Wolt
et al. 2002) major radio ejections (which were directly resolved)
followed the plateau. Thus, as in the other sources, a transition from
the 'hard' to the 'soft' VHS/IS (as GRS 1915+105 rarely, if ever,
reaches the canonical LS or HS) resulted in a major ejection. However,
the data also reveal that the radio spectrum started to change {\em
prior} to the first major ejection event. This is presented in Fig 2.

In fact the radio spectrum is already 'optically thin' a couple of
days before the flux starts rising, reaching a mean value of -0.6 on
MJD 50750, the day before the flare. The change in the radio spectrum
occurs around the time that the spectrum starts to soften dramatically
towards the 'soft' VHS/IS.

\subsubsection{GX 339-4}

The detailed radio light curve of GX 339-4 around the time of the
major radio flare is reported in Gallo et al. (2004).  During the
first 2 hours of observations on MJD 52408, immediately prior to the
rise of the radio flare, the spectral index seems to vary from flat to
more optically thin ($\alpha \sim -0.2$) and back to flat. Even though
the error bars are relatively large, this is consistent with rapid
variability of the radio spectrum outflow structure prior to the radio
flare.

\subsubsection{XTE J1859+226}

In this source, over the 4 days prior to the flare on MJD 51467.5, the
radio spectrum changes from flat ($0.037 \geq \alpha \geq -0.030$) to
more optically thin ($\alpha=-0.237$ on MJD=51467); see Fig 2 (lower
panels). On this date, the radio spectrum is measured over seven
frequencies: fig. 5 of Brocksopp et al. (2001) shows that the actual
spectrum, even though overall optically thin, is peaked around 3 GHz
(i.e. flat-inverted till ~3 GHz and optically thin above), indicating
that a structural change in the outflow has occurred prior to the
flare. 

\subsubsection{XTE J1550-564}

The radio coverage of the outburst of XTE J1550-564 as presented in
Fig 1(d) is reported in Wu et al. (2002). However, further
measurements presented in Hannikainen et al. (2004) indicate that on
MJD 51073, some 4.5 days prior to the flare, there was rapid spectral
variability. This included a transition from a peaked to an optically
thin spectrum, and back to a peaked spectrum again, on a timescale of
less than half a day.

\subsection{Association of the outburst with the soft VHS/IS peak}

Many -- perhaps all -- sources which undergo an X-ray outburst have a
bright hard state during the rising phase (e.g. Brocksopp et al. 2002;
Maccarone \& Coppi 2003; Yu et al. 2003; Yu, van der Klis \& Fender
2004). We know that hard X-ray states correspond to phases of powerful
jet production, and that the relativistic ejections tend to occur
around the time of X-ray state transitions, and that soft states do
not seem to produce jets (see Fender 2004 for a review).

What was difficult to tell from investigation of the most-studied
sources like GRS 1915+105 and GX 339-4, was to what part of the state
transition sequence the optically thin jet formation corresponded. The
above examples appear to give a clear answer -- the optically thin
radio jet is produced at the VHS/IS peak, which occurs at or near the
end of the X-ray spectral softening.  We note that this is consistent with
the suggestion of Mirabel et al. (1998) that the 'spike' in the X-ray
light curve of GRS 1915+105 corresponded to the trigger point for the
optically thin radio event although, as we have seen, in the case of
GRS 1915+105 there are many ambiguities which cannot be resolved by
studying this source in isolation.

Figure 3 presents the hardness-intensity diagrams (HIDs) for the three
of the four sources presented in Fig 1, with the point corresponding
most clearly to the time of the major radio flare indicated with a
circle. In these HIDs the canonical LS corresponds to a nearly
vertical branch on the right hand side of the diagram, seen here only
for GX 339-4; the horizontal right-to-left motion exhibited by these
sources is characteristic of a transition from a 'hard' to a 'soft'
VHS/IS (Belloni 2004; Belloni et al. 2004; Homan \& Belloni
2004). Furthermore, the X-ray intensity has been converted to an
Eddington-fraction luminosity, based upon the distance and mass
estimates given in table 1.  It is clear that in all cases the radio
flare occurs when the sources are in the VHS/IS, having significantly
softened from the LS prior to the event, and with a X-ray luminosity
in the range 10--100\% Eddington.

It is worth revisiting the result of Corbel et al. (2001) in which XTE
J1550-564 was found to be radio-quiet while in the VHS/IS. An
inspection of the X-ray data, indicates the following: the first radio
observation, which resulted in an optically thin detection, was made
within a day or two of the soft VHS/IS peak.  The second observation,
resulting in an upper limit only, was several days late, in the middle
of the soft VHS/IS phase.  Thus these observations are consistent with
the picture presented here, namely that the optically thin radio flare
occurs at the soft VHS/IS peak, and in the subsequent soft VHS/IS
phase the core radio emission is suppressed.

Corbel et al. (2004) report a detailed study of the radio emission
from the black hole transient XTE J1650-500 (plus a phenomenological
comparison with XTE J1859+226 and XTE J1550-564) in which the
canonical LS and HS, as well as various 'degrees' of VHS/IS were
observed. Specifically they conclude that the steady LS jet may
persist into the 'hard' VHS/IS, and that the optically thin radio
flare is associated with the soft VHS/IS peak (although they use the
state definitions of McClintock \& Remillard 2004). These results seem
to be consistent with the conclusions we have drawn (more
quantitatively) about the association of X-ray states and jet
production. 

\subsection{The variation of accretion disc radius with state}

It is widely accepted that a geometrically thin, optically thick,
accretion disc extends close to the black hole in 'soft' X-ray states
and is 'truncated' at larger radii in 'harder' X-ray states
(e.g. Esin, McClintock \& Narayan 1997; McClintock \& Remillard 2004
and references therein).  While the absolute values of the radii
obtained from X-ray spectral fits may be severely underestimated
(e.g. Merloni, Fabian \& Ross 2000), large changes in the fitted radii
are likely to be due to real changes in the location of the brightest
X-ray emitting region. Specifically, for the four sources under
detailed consideration here:

\subsubsection{GRS 1915+105}

The 'dip-flare' cycles of GRS 1915+105, such as those presented in Fig
1(a), are well known to be associated with apparent changes in the
fitted accretion disc radius (e.g. Belloni et al. 1997; Fender \&
Belloni 2004 and references therein). During the 'soft VHS/IS' (states
A/B) the fitted inner disc radius reaches a stable, low, value and is
considerably larger in the 'hard VHS' (state C).

\subsubsection{GX 339-4}

Spectral fits over the period focussed on in Fig 1(b) indicate a
fitted inner disc radius which decreased dramatically at the point of
spectral softening (Zdziarski et al. 2004; Nespoli 2004). This low value of the
fitted inner radius remained stable for an extended period ($> 100$
days) until the return to the canonical LS.

\subsubsection{XTE J1859+226}

To our knowledge, detailed spectral fits over the entire outburst of
XTE J1859+226 have not been published. However, both the general X-ray
spectral and timing evolution and the preliminary fits reported by
Markwardt (2001) indicate a similar pattern to other X-ray binaries.
Note that Hynes et al. (2002) discuss the post-outburst evolution of the
accretion disc in XTE J1859+226 but note that the absolute value of
the inner disc radius cannot be well constrained and so it is hard to
use their results to compare with earlier in the outburst.

\subsubsection{XTE J1550-564}

Sobczak et al. (2000) fitted disc radii to over 200 spectra of XTE
J1550-564 during the entire 1998--1999 outburst. They found that at
the peak of the outburst ('soft VHS') the fitted disc radii varied a
lot but were in general very (unrealistically) small. Subsequently in
the canonical HS state the disc radius remained relatively
small and stable over $\sim 100$ days.

\subsection{The alternative state definitions of McClintock \&
  Remillard}

We can summarise these connections between X-ray state and radio
emission within the framework of the revised definitions of McClintock
\& Remillard (2004). This is presented in Table 2.

\begin{table}
\begin{tabular}{ccc}
\hline
'Classical' & McClintock \& & Radio properties \\
states      & Remillard  & \\
\hline
Low/Hard state (LS) & Hard state & steady jet \\
'Hard' VHS/IS & Intermediate state & steady jet \\
'Hard' $\rightarrow$ 'soft' VHS/IS &  Intermediate
$\rightarrow$ SPL & radio flare \\
'Soft' VHS/IS & Intermediate / SPL & no jet \\
High/Soft state (HS) & Thermal dominant & no jet \\
\hline
\end{tabular}
\caption{Comparison of the 'classical' X-ray states, and those of
  McClintock \& Remillard (2004), with the radio properties as
  discussed in detail in the text. 'VHS/IS' corresponds to 'Very High
  State/Intermediate State' and 'SPL' stands for 'Steep Power Law'.} 
\end{table}

Two significant points are worth noting:

\begin{enumerate}
\item{In the framework of McClintock \& Remillard, it may be exactly
  at the point of the transition to the SPL that corresponds to the
  radio ejection event.}
\item{In the same framework, there appear to be both 'jet on' and 'jet
  off' phases associated with the same state 'intermediate' label.}
\end{enumerate}

Point (i), above, was already suggested by McClintock \& Remillard
(2004) has been noted as likely also by Corbel et al. (2004). However,
considering point (ii), it appears that the new definitions have both
advantages and disadvantages.

\section{Increasing velocity as a function of X-ray luminosity at
  launch}

\begin{figure}
\label{figvelx}
\centerline{\epsfig{file=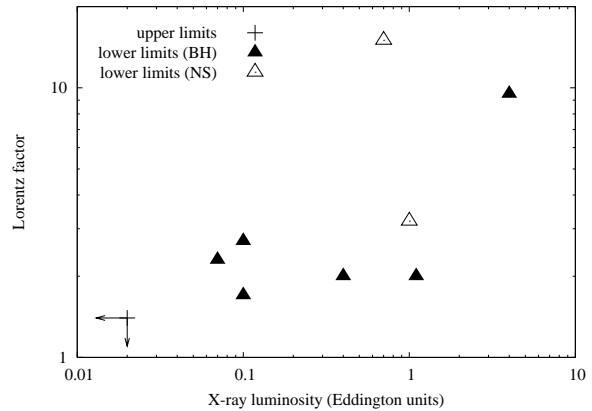, angle=270, width=8cm}}
\caption{Limits on jet Lorentz factors as a function of the estimated
  bolometric X-ray luminosity at the time of jet launch. The arrows in
  the lower left of the figure indicate the condition that jets in the
  general LS state have $v \leq 0.8c$. The rest of the symbols
  are lower limits only to the Lorentz factors from individual black
  hole (filled symbols) and neutron star (open symbols) X-ray
  binaries; the data are listed in table 1.}
\end{figure}

A further key component for the model towards which we are progressing
is the variation of jet velocity with X-ray luminosity / state.  The
empirical evidence clearly points to an increase of jet velocity with
X-ray luminosity, at least in the sense of a step up from mildly
relativistic velocities in the LS to significantly relativistic
velocities resulting from outbursts at a significant fraction of the
Eddington limit. Table 1 presents a compilation of estimated jet
Lorentz factors as a function of the X-ray luminosity at launch. In
fact, with the exception of the upper limit for the LS sources (see
below), these are all based upon observed proper motions in spatially
resolved radio maps and are therefore only lower limits (see Fender
2003).  These data are plotted in Fig {\ref{figvelx}}. Note that a
higher Lorentz factor in transient jets associated with outbursts is
further supported by the much larger scatter in the $L_{\rm
radio}:L_{\rm X}$ plane compared to the LS sources (Gallo, Fender \&
Pooley 2003).

The measured spread to the radio/X-ray correlation in LS black hole
X-ray binaries was interpreted by Gallo, Fender \& Pooley (2003) in
terms of a distribution in Doppler factors and used to infer an upper
limit $\Gamma \la 2$ to the bulk velocity of compact jets. The value
of the spread as it appears in GFP03 is mainly determined by the two
data sets for which the correlation extends over more than three
orders of magnitude in X-ray luminosity, namely V 404 Cyg and GX
339-4.  Obviously, errors in the distance estimates to these systems
will introduce a further source of scatter to the correlation. While V
404 Cyg is (reasonably well) known to lie at $\sim$3.5 kpc (Wagner
et. al 1992), the distance to GX 339-4, and hence its luminosity,
remain uncertain. Recent works (Hynes et al 2004; Zdziarski et
al. 2004) have significantly revised the lower limit of 4 kpc
(Zdziarski et al. 1998) adopted in GFP03, placing GX 339-4 at a
distance larger than 6 kpc. In particular, Hynes et al. (2004) discuss
the possibility that the system could even be located on the far side
of the Galaxy, at $\sim$ 15 kpc. In fact, any value between 6 and 15
kpc has the effect of lowering the final spread to the radio/X-ray
correlation by shifting GX 339-4 closer to V 404 Cyg, resulting in a
more stringent upper limit on the jet bulk Lorentz factor. A minimum
distance to GX 339-4 of 6 kpc reduces the measured spread by a factor
1.6, requiring an outflow bulk velocity smaller than 0.7$c$. This sets
a new upper limit of $\Gamma_{radio}\la 1.4$ to the average bulk
Lorentz factor of LS black hole X-ray binary jets; in fact, the
distance to GX 339-4 would have to be larger than 15 kpc before the
spread exceeds again the value obtained for 4 kpc.  As discussed in
GFP03, the above arguments are formally valid in case of radio beaming
combined with isotropic X-ray emission, but if X-rays are moderately
beamed ($v_{X}\sim 0.3$c, as suggested by e.g. models of dynamical
coronae, Beloborodov 1999), the conclusions remain essentially
unchanged.

However, Heinz \& Merloni (2004) argue that the spread around the
correlation can only really be used, in the absence of additional
information, to constrain the {\em range} in jet velocities, and not
the absolute values. Based upon their arguments, it remains likely
that the bulk Lorentz factor of the LS jets is $\Gamma \sim 1$ but it
is not a formal requirement. Based upon an analysis of the
normalisations for GX 339-4 and V404 as estimated in Gallo, Fender \&
Pooley (2003) they conclude that the Lorentz factors of the two
sources differ by no more than a factor of two. Given the
'universality' -- within about one order of magnitude -- of the
correlation presented in Gallo, Fender \& Pooley (2003), this in turn
implies that independent measurement of the Lorentz factor of a jet in
the LS would apply to all LS sources. In the discussion that follows
we shall continue to assume that the Lorentz factor of the steady jets
$\Gamma \leq 2$ but note that it is not proven.

What is not clear from these data is whether the velocity is a simple
`step' function of the X-ray luminosity at launch, or rather a
smoother function. As discussed in Fender (2003) a broad range in
$\Gamma$ will, in most circumstances, produce approximately the same
proper motions.  For the simple unified model discussed in this
article either ('step' or smooth function) interpretation is
sufficient. The key factor is that the jets associated with the VHS/IS
peak are probably more relativistic than those in the LS which
generally precedes it.

\section{Radio emission and jet power}

It is also crucial to estimate the jet power as a function of X-ray
luminosity / state. In the following we present simplified expressions
for the power in both optically thick and optically thin jets, in
Eddington units, as a function of observable radio and X-ray emission.

\subsection{The low/hard state optically thick jet}

In Fender, Gallo \& Jonker (2003) it was argued that the total jet power
$L_{\rm J}$, in the absence of significant advection, was related to
the accretion luminosity $L_{\rm X}$ as follows:

\[
L_{\rm J} = A_{\rm steady} L_{\rm X}^{0.5}
\]

where $A_{\rm steady} \geq 6 \times 10^{-3}$ (the normalisation is
referred to simply as $A$ in Fender, Gallo \& Jonker 2003).

Studies of the rapid variability from the 'hard' transient XTE
J1118+480, which remained in the LS throughout its outburst, have
supported the idea that the optical emission may originate in an
outflow and not reprocessed emission from the disc (Merloni, di Matteo
\& Fabian 2000; Kanbach et al. 2001; Spruit \& Kanbach 2002; Malzac et
al. 2004). Detailed modelling of the correlated variability by Malzac,
Merloni \& Fabian (2004) has resulted in a normalisation of the
jet/outflow power which corresponds to $A_{\rm steady} \sim 0.3$ in the above
formalisation, which would imply that all LS sources are
jet-dominated. For now we shall take this as the largest likely value
of $A_{\rm steady}$ (see also Yuan, Cui \& Narayan 2004 who estimate a
value for the radiative efficiency for the jet in XTE J1118+480 which
lies between the lower limit of Fender, Gallo \& Jonker 2003 and the
estimate of Malzac, Merloni \& Fabian 2004).

\subsection{The optically thin jets}

The power associated with the production of optically thin jets can be
calculated from the peak luminosity and rise time of the event,
adapting the minimum energy arguments of Burbidge (1956, 1959), as follows:

\[
L_{\rm J} = 20 \Delta t^{2/7} L_{\rm radio}^{4/7} M^{-3/7} = 2
\times 10^{-5} \Delta t^{2/7} d^{8/7} S_{\rm 5GHz}^{4/7} M^{-1}
\]

where $L_{\rm J}$ is the mean power into jet production (in Eddington
units), $\Delta t$ is the rise time of the event, in seconds, $L_{\rm
radio}$ is the peak radio luminosity of the event at 5 GHz (in
Eddington units), $d$ is the distance in kpc, S$_{\rm 5GHz}$ is the
peak radio flux density at 5GHz (in mJy), and $M$ is the black hole
mass in solar units.  The equation assumes an emitting plasma with
volume corresponding to $4 \pi (\Delta t c)^3$, a filling factor of unity,
negligible energy in protons and a spectral index of $\alpha =
-0.75$. See Longair (1994) for a fuller discussion.

In addition, since we have argued above that the bulk Lorentz factor
is considerably higher for the transient jets underlying these
optically thin outbursts, we need to compensate for the resultant
Doppler effects. Fender (2001b) demonstrated that it is much more
likely, for significantly relativistic jets, that the jet power is
underestimated than overestimated, and introduced a correction factor
$F(\Gamma, i) = \Gamma \delta^{-3}$ where $\delta$ is the relativistic
Doppler factor associated with bulk Lorentz factor $\Gamma$ (the
correction includes the kinetic energy of bulk flow). For $2
\leq \Gamma \leq 5$ the mean value of $F(\Gamma, i)$ averaged over
$\cos (i)$ is $\sim 50$.
We adopt this value as an additional
(upward) correction to the power of the optically thin jets. Comparison of the
formula given above with specific examples more carefully considered,
e.g. the 1997 ejections from GRS 1915+105 reported by Fender et
al. (1999), indicate this to be a reasonable correction. As discussed
earlier, however, we have no clear upper limit on $\Gamma$ associated
with these events, and the correction could be much larger. For
example, for $2 \leq \Gamma \leq 7$ the mean value of $F(\Gamma, i)$
is $\sim 160$, and for $2 \leq \Gamma \leq 10$ is it $\sim 575$.

In table 1 we list in columns eight and nine the estimated optically
thin radio powers during the flare events, $L_{\rm J}$, and the
corresponding peak 'soft VHS/IS' X-ray luminosity, $L_{\rm X, VHS}$,
both expressed as Eddington fractions. These values are plotted
against each other in Fig 5, and compared with the functions for the
steady / LS jets as outlined above, for both the Fender, Gallo \&
Jonker (2003) lower limit and the Malzac, Merloni \& Fabian (2004)
estimate.

A best-fit power-law
to the data for the transient events is of the form

\[
L_{\rm jet} = A_{\rm trans} L_{\rm X}^{0.5 \pm 0.2}
\]

where the fitted value is $A_{\rm trans} = (0.4 \pm 0.1)$. Note that
since for the transient jets $L_{\rm X} \sim 1$ this indicates near
equipartition of $L_{\rm X}$ and $L_{\rm J}$ around the time of such events.

The index of the fit, $0.5 \pm 0.2$ is comparable to that derived for
the LS, namely $L_{\rm jet} \propto L_{\rm X}^{0.5}$ (Fender, Gallo \&
Jonker 2003).  The value of the normalisation, $A_{\rm trana}$, is
much larger than the conservative value for the steady jet
normalisation $A_{\rm steady}$ estimated in Fender, Gallo \& Jonker
(2003; see above). However, it is only $\sim 50$\% larger than the
value of $A_{\rm steady}$ implied by the results of Malzac, Merloni \&
Fabian (2004).  Were such a large value to be valid for the LS it
would imply that black hole X-ray binaries are likely to be
jet-dominated below $L_{\rm X} = A_{\rm steady} ^2 \sim 0.1 L_{\rm
Edd}$. Since most sources do not strongly exceed this Eddington ratio
while in a 'hard' X-ray state (see e.g. Figs 3 and 5), it implies that
{\em only} in the HS and 'soft' VHS/IS and states, when the jet is
suppressed, is the X-ray luminosity dominant over the jet (see also
discussion in Malzac et al. 2004). Put another way, for such a large
normalisation, whenever the jet is on, it is the dominant power output
channel.

\begin{figure}
\label{softcorr}
\centerline{\epsfig{file=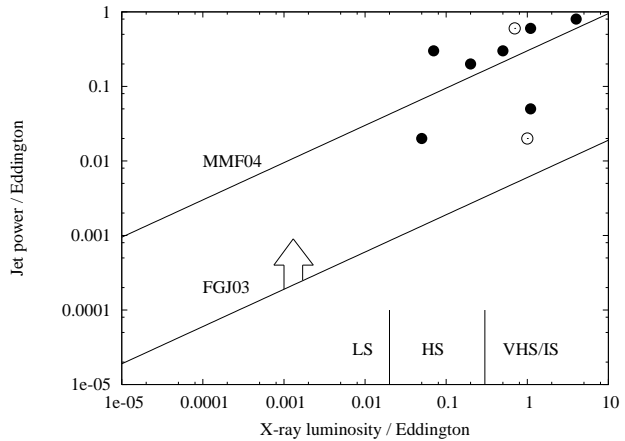, angle=270, width=8.5cm}}
\caption{Comparing estimated jet power in the steady / LS and
  transients / VHS/IS states. Note that the uncertainties on the
  estimated jet powers are large, {\em at least} one order of
  magnitude. The solid line marked FGJ03 indicates
  the lower limit to the steady jet power in the LS as estimated by
  Fender, Gallo \& Jonker (2003). The line marked MMF04 indicates the
  same function with the larger normalisation as calculated by Malzac,
  Merloni \& Fabian (2004). The (solid) points indicate the estimates of jet
  power for the transient events listed in table 1. It is interesting
  to note that the data are therefore compatible with the MMF04
  relation at all X-ray luminosities. If the steady LS jet power is
  lower, nearer to the lower limit indicated by the FGJ03 function,
  then there may be a step up in jet power for the transient
  events. A very approximate indication of the regimes typically
  corresponding to the different X-ray states is indicated at the
  bottom of the figure. The power estimates for the two neutron star
  sources, Sco X-1 and Cir X-1, are indicated by open symbols.}
\end{figure}

\subsection{Some caveats}

Note that there are very large uncertainties remaining in the
estimation of both $A_{\rm steady}$ and $A_{\rm trans}$. The functions
used for the power of the steady jets are based upon essentially one
detailed example -- XTE J1118+480 -- and extrapolated to other sources
via the 'universal' $L_{\rm radio} \propto L_{\rm X}^{0.7}$ relation
(although estimates from a handful of other sources are also
compatible).  While this approach may well be appropriate, spectral
changes -- in particular the location of the optically thin break in
the synchrotron spectrum -- could strongly affect the variation of
$L_{\rm J}$ as a function of $L_{\rm X}$ and more work needs to be
done in the future.

In the case of the power function for the transient jets, we have in
fact more independent measurements, but those measurements themselves
probably have a greater associated uncertainty.  Underestimation of
$L_{\rm J}$ can clearly arise due to deviations from equipartition, a
lack of knowledge of the high-frequency extension of the synchrotron
spectrum and a possible underestimate of the bulk Lorentz
factor. Overestimation of $L_{rm J}$ could arise due to overestimation
of the synchrotron-emitting volume (ie. a filling factor $f < 1$ or
injection/acceleration of particles into a confined jet).

\subsection*{A single function or a step up in jet power ?}

Nevertheless, it is noteable that a single power-law relation could be
plotted through both the steady LS and transient VHS/IS functions, and
that the normalisation of such a single function would be close to
that estimated by Malzac, Merloni \& Fabian (2004), i.e. $A_{\rm
  steady} \sim A_{\rm trans} \sim 0.3$.

If there is not a single function then it seems that the transient
VHS/IS jets are somewhat more powerful as a function of $L_{\rm X}$
than an extrapolation of the steady LS jets function. This suggestion
is strengthened by our argument, below, that the optically thin events
are likely to arise in internal shocks which do not dissipate 100\% of
the available kinetic energy.  If real, this effect may be due to
temporarily increased efficiency of jet production in the inner disc,
or the transient addition of a new power source, namely the black hole
spin (e.g. Blandford \& Znajek 1977; Punsly \& Corotoni 1991; Livio,
Ogilvie \& Pringle 1999; Meier 1999, 2001, 2003; Koide et
al. 2002). Nevertheless, we consider the similarity in both gradient
and normalisation of the two jet functions to be remarkable. We note
that for the model outlined in this paper to be tested against
higher-mass (intermediate or supermassive) black holes, a further mass
term would be required for both expressions (see Heinz \& Sunyaev
2003; Merloni, Heinz \& di Matteo 2003; Falcke, K\"ording \& Markoff
2004). However, for the X-ray binaries where the range in mass is
likely to be $\leq 2$ this is not important at the current level of
accuracy.

\section{Internal shocks}

The arguments given above clearly indicate that as the X-ray
luminosity of the accreting source increases, then so does the
velocity of the outflow (although whether this is in the form of a
step, or other functional form, is as yet unclear). Since most,
probably all, outbursting sources have followed a path in which they
have become monotonically brighter in a hard state before making a
transition to a soft state, this tells us that a shock should form in
the previously-generated `steady' jet as the faster-moving VHS/IS jet
catches up and interacts with it. This internal shock is therefore a
natural origin for the optically thin events observed at the beginning
of X-ray transient outbursts. Internal shocks have previously been
proposed for AGN (e.g. Rees 1978; Marscher \& Gear 1985; Ghisellini
1999; Spada et al. 2001) and gamma-ray bursts (GRBs) (e.g. Rees \&
Meszaros 1994; van Paradijs, Kouveliotou \& Wijers 2000 and references
therein).  Indeed in the context of X-ray binaries an internal-shock
scenario has already been discussed previously for GRS 1915+105 by
Kaiser, Sunyaev \& Spruit(2000), Vadawale et al. (2003) and Turler et
al. (2004), and their ideas have significantly inspired this work.

Rees \& Meszaros (1994) spelled out the basis for such 'internal
shocks' in the context of GRBs. They assumed two
'blobs' of equal mass but differing Lorentz factors were ejected such
that the later ejection had the higher Lorentz factor. This component,
if moving along precisely the same trajectory as the original
component, will collide with it. Assuming conservation of energy and
momentum it was shown that up to 40\% of the total kinetic energy
could be released in this shock.

The formulation for the efficiency of energy release, $\epsilon$, as
presented in Rees \& Meszaros corresponds to the case in which both
blobs have Lorentz factors $\Gamma_{1,2} >> 1$. We have repeated their
approach, but considered instead the (less simple) case in which the
first blob is at most only mildly relativistic ($\Gamma_1 < 2$).

In Fig 7 we plot the internal shock efficiency for two cases:

\begin{itemize}
\item{{\bf (a)} The blobs have the same total energy, fulfilling the criterion
  that $\Gamma_1 M_1 = \Gamma_2 M_2$ (where $M$ is the mass
  of the blob). This corresponds to the situation in
  which the jet power does not increase significantly, while the
  Lorentz factor does.}
\item{{\bf (b)} The blobs have equal mass. This corresponds to a genuine
  increase in jet power by a factor $\Gamma_2 - \Gamma_1 \sim
  \Gamma_2$. This corresponds to the maximum efficiency for the
  internal shock (e.g. Spada et al. 2001).}
\end{itemize}

For the kind of values estimated for transient jets, efficiencies in
the range 5--40\% are expected.  For a total jet power of $P_{\rm J}$,
the power in the shock (i.e. available for particle acceleration) will
be $P_{\rm shock} \leq \epsilon P_{\rm J}$.  The remaining jet power
$(\epsilon - 1)P_{\rm J}$ is associated with the kinetic energy of the
merged shells and should eventually be dissipated via interactions
with the ambient medium.

\begin{figure}
\centerline{\epsfig{file=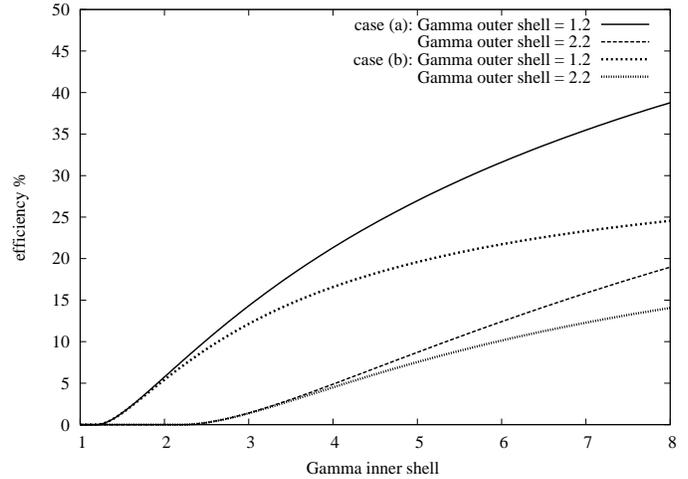, width=6.5cm, angle=270}}
\caption{Efficiency of energy release for the collision of two shells,
  where the outer shell is only mildly relativistic (bulk Lorentz
  factor [BLF] $\Gamma<2$, under the assumption of conservation of
  momentum and energy. Case (a) corresponds to the case of
  two blobs with equal total energy (i.e. $\Gamma_1 M_1 = \Gamma_2
  M_2$) in which case the jet power has not increased. Case
  (b) corresponds to the case of the same mass but increased Lorentz
  factor, ie. an increase in jet power by a factor $\Gamma_2 -
  \Gamma_1$.  This can be considered a (high oversimplified)
  approximation to the collision of a relativistic VHS/IS jet (inner
  shell) with the
  preceding steady LS jet (outer shell).  }
\end{figure}

Of course this is a highly oversimplified approximation of the true
circumstances as the jet increases in velocity.  Nevertheless, it
illustrates that the internal shock produced by a transient (as it is
subsequently shut down in the soft state) acceleration of the jet from
$\Gamma \sim 1$ to $\Gamma \la 10$ could produce dissipation in an
internal shock with an observed energy release comparable (within the
order or magnitude uncertainty) to that estimated for the steady jet
prior to the acceleration. If, as seems likely, $A_{\rm trans} \geq
A_{\rm steady}$ then the more efficient shock scenario (b) (Fig 7) is
more likely, and the total jet power, and not just velocity, has
significantly increased. Beloborodov (2000) discusses in further
detail the high radiative efficiencies which may be obtained in the
internal shock model.

The internal shock scenario is also attractive as an explanation for
why the same radio flux at a given radio frequency for a given source
can be sometimes optically thin and sometimes optically
thick. Consider GRS 1915+105, where in the plateau states a flux
density of $\sim 40$ mJy at 15 GHz may be associated with an optically
thick spectrum, and later a comparable flux density with optically thin
rising phases of oscillation events (e.g. Fender et al. 1999). If the
particle acceleration all occurred at the base of a jet with an
approximately fixed structure, this is hard to explain. However, it
follows naturally from a scenario where the optically thin events are
associated with internal shocks occurring at a much larger distance
from the dense inner jet.

Note that it is the radiation resulting from the energy liberated by
the internal shock, which we have measured in order to estimate the
jet power in section 5 above. However, since our estimates of the bulk
Lorentz factor must be based upon observations of the post-shock
plasma, then the true jet power must be larger by a factor
$\epsilon^{-1}$. Since $0.05 \la \epsilon \la 0.45$ in the above
simplification, this may imply that the underlying jet power is
actually a further order of magnitude larger for the transient
jets. In this case a single function corresponding to both the LS and
VHS/IS jets seems less likely.

As discussed in Vadawale et al. (2003) the strength of the shock is
likely to be related to the amount of material lying in the path of
the faster 'VHS/IS' jet. They discussed this in the context of GRS
1915+105, where the strength of 'post-plateau jets' (Klein-Wolt et
al. 2002) is shown to be correlated with the total X-ray fluence of
the preceding 'plateau' (which was presumably a phase of slower jet
production). Generalising this phenomenon to other X-ray transients,
it provides a natural explanation for why, although there are often
multiple radio flaring events, the first is invariably the strongest.

\begin{figure*}
\label{toymodel}
\centerline{\epsfig{file=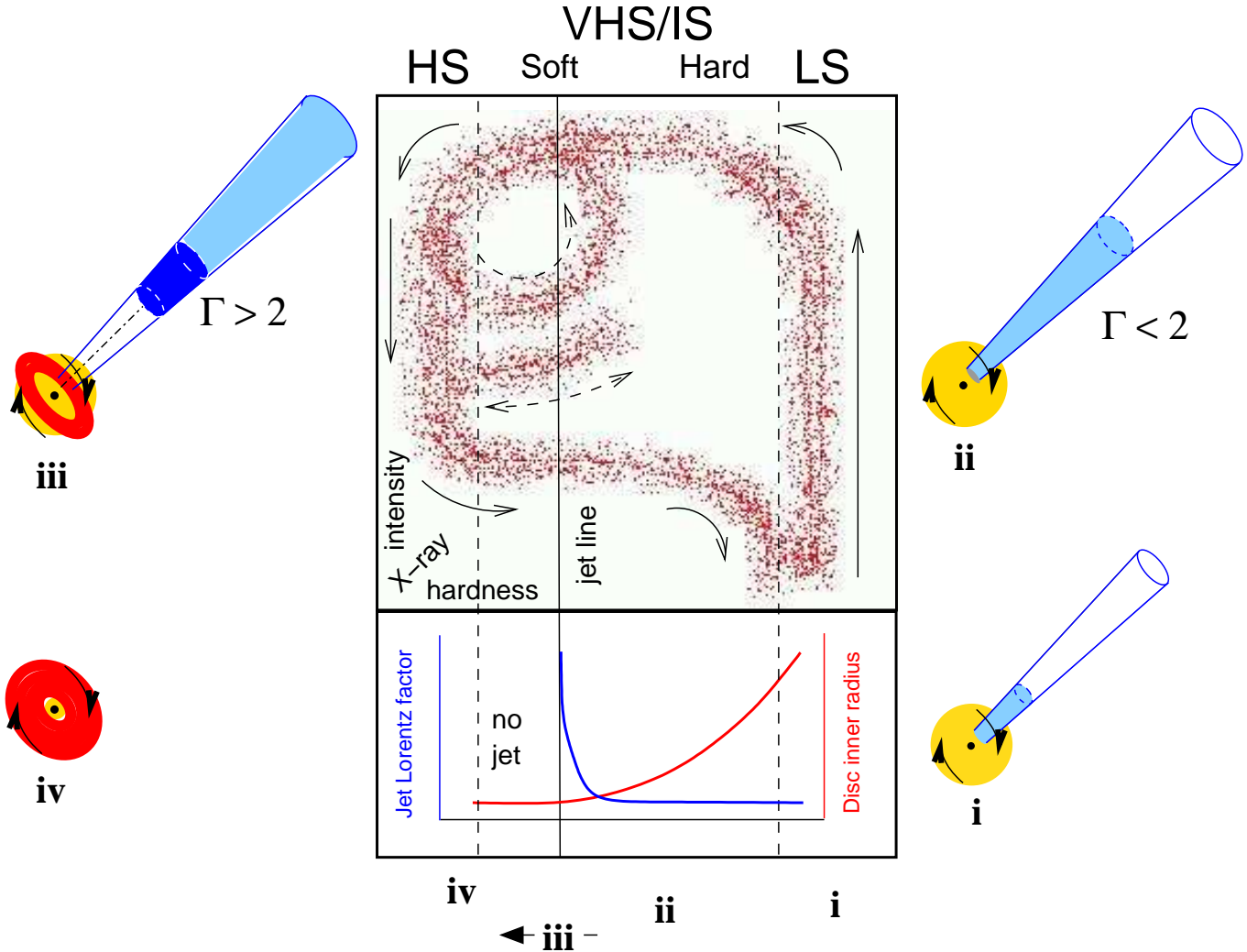, angle=0, width=18cm}}
\caption{A schematic of our simplified model for the jet-disc coupling
  in black hole binaries. The central box panel represents an X-ray
  hardness-intensity diagram (HID); 'HS' indicates the `high/soft
  state', 'VHS/IS' indicates the 'very high/intermediate state' and
  'LS' the 'low/hard state'. In this diagram, X-ray hardness increases
  to the right and intensity upwards. The lower panel indicates the
  variation of the bulk Lorentz factor of the outflow with hardness --
  in the LS and hard-VHS/IS the jet is steady with an almost constant
  bulk Lorentz factor $\Gamma < 2$, progressing from state {\bf i} to
  state {\bf ii} as the luminosity increases. At some point -- usually
  corresponding to the peak of the VHS/IS -- $\Gamma$ increases
  rapidly producing an internal shock in the outflow ({\bf iii})
  followed in general by cessation of jet production in a
  disc-dominated HS ({\bf iv}). At this stage fading optically thin
  radio emission is only associated with a jet/shock which is now
  physically decoupled from the central engine.  As a result the solid
  arrows indicate the track of a simple X-ray transient outburst with
  a single optically thin jet production episode. The dashed loop and
  dotted track indicate the paths that GRS 1915+105 and some other
  transients take in repeatedly hardening and then crossing zone {\bf
  iii} -- the 'jet line' -- from left to right, producing further optically thin radio
  outbursts. Sketches around the outside illustrate our concept of the
  relative contributions of jet (blue), 'corona' (yellow) and
  accretion disc (red) at these different stages.}
\end{figure*}

\section{Towards a unified model}

Based upon the key generic observational details assembled above, we
have attempted to construct a unified, semi-quantitative, model for
the disc-jet coupling in black hole X-ray binaries. A simplified
version of the model specific to GRS 1915+105 has been presented in
Fender \& Belloni (2004).  The model is summarised in Fig 7, which we
describe in detail below. The diagram consists of a schematic X-ray
hardness-intensity diagram (HID) above a schematic indicating the bulk
Lorentz factor of the jet and inner accretion disc radius as a
function of X-ray hardness. The four sketches around the outside of
the schematics indicate our suggestions as to the state of the source
at the various phases {\bf i}--{\bf iv}. The path of a typical X-ray
transient is as indicated by the solid arrows.

\begin{itemize}
\item{Phase {\bf i}: Sources are in the low-luminosity LS, producing a
  steady jet whose power correlates as $L_{\rm jet} \propto L_{\rm
  X}^{0.5}$ (ignoring any mass term). This phase probably extends down
  to very low luminosities ('quiescence')}.
\item{Phase {\bf ii}: The motion in the HID, for a typical outburst,
  has been nearly vertical. There is a peak in the LS after which the
  motion in the HID becomes more horizontal (to the left) and the
  source moves into the 'hard' VHS/IS. Despite this softening of the
  X-ray spectrum the steady jet persists, with a very similar
  coupling, quantitatively, to that seen in the LS.}
\item{Phase {\bf iii}: The source approaches the 'jet line' (the
  solid vertical line in the schematic HID) in the HID between jet-producing
  and jet-free states. As the boundary is approached the jet
  properties change, most notably its velocity. The final, most
  powerful, jet, has the highest Lorentz factor, causing the
  propagation of an internal shock through the slower-moving outflow
  in front of it.}
\item{Phase {\bf iv}: The source is in the 'soft' VHS/IS or the
  canonical HS, and no jet is produced. For a while following the peak
  of phase iii fading optically thin emission is observed from the
  optically thin shock.}
\end{itemize}

Following phase {\bf iv}, most sources drop in intensity in the
canonical HS until a (horizontal) transition back, via the VHS/IS, to
the LS. Some sources will make repeated excursions, such as the loops
and branches indicated with dashed lines in Fig 7, back across the jet
line, However, with the exception of GRS 1915+105, the number of such
excursions is generally $\leq 10$.  When crossing the jet line from
right to left, the jet is re-activated but there is (generally) no
slower-moving jet in front of it for a shock to be formed; only motion
from left to right produces an optically thin flare (this is a
prediction). Subsequently the motion back towards quiescence is almost
vertically downwards in the HID.

The model as outlined above has many similarities with the scenarios
described by Meier (1999, 2001, 2003) who has approached the problem
from a more theoretical point of view. Meier (2001) has suggested that
in low-luminosity states the jet is powered by a modification of the
Blandford \& Payne ('BP') (1982) mechanism taking into account frame-dragging
near a rotating black hole (Punsly \& Coroniti 1990). This 'BP/PC
mechanism' can extract black hole spin by the coupling of magnetic
field lines extending from within the ergosphere to outside of
it. Meier (2001) further suggests that during phases of very high
accretion the Blandford \& Znajek ('BZ') (1977) mechanism may work
briefly. This may be associated with a 'spine jet' which is
considerably more relativistic than the 'sheath jet' produced by the
BP/PC mechanism. Note that the power of the jets as given in Meier
(2001, 2003) is about linearly proportional to the accretion rate; in
the formulation of Fender, Gallo \& Jonker (2003) this corresponds to
the 'jet dominated state' (see also Falcke, Kording \& Markoff 2004).

We can revisit the scenarios of Meier in the light of our compilation
of observational results and steps toward a unified model. In the
faint LS (phase {\bf i} in Fig 7) is the jet formed by the BP or BP/PC
mechanisms ?  Given that the jet may be formed at relatively large
distances from the black hole in such states, there may not be any
significant influence of the black hole spin on the jet formation
process. However, it is also likely that in such states the
jet-formation process is not occurring within thin discs, as is the
basis of the BP mechanism, but rather in a geometrically thick flow
(see also e.g. Blandford \& Begelman 1999; Meier 2001; Merloni \&
Fabian 2002)

As the accretion rate increases the power of this disc-jet will
increase and the geometrically thin accretion disc will propagate
inwards. During this phase the jet formation process may migrate from
BP$\rightarrow$BP/PC. However, the suggestion that the most
relativistic jets are formed by the BZ process seems at odds with the
observation of significantly relativistic outflows from two neutron
stars systems (Fomalont et al. 2001a,b; Fender et al. 2004).  In a
related work, the results of Yu, van der Klis \& Fender (2004)
indicate that the subsequent evolution of X-ray transient outbursts is
approximately determined {\em before} the soft VHS/IS peak, in both
neutron star and black hole systems. This suggests that already by
the time of the LS peak we can estimated the size of the ejection even
which is to follow, and is a further indication that the study of neutron
stars will shed important light on the physics of jet formation in
black hole systems.

\subsection{Ejected disc or ejected corona ?}

It is interesting to compare the sequence of events we have outlined
as being responsible for ejection events with the interpretation most
commonly put forward when the disc-jet coupling in GRS 1915+105 was
first observed. In this source oscillations, on timescales of tens of
minutes, between hard (state 'C' $\equiv$ hard VHS/IS) and soft
(states 'A' and 'B' $\equiv$ soft VHS/IS -- see Fig 1(a)) were
associated with cycles of disc 'emptying' and refill (e.g. Belloni et
al. 1997b). When the relation to ejection events (Pooley \& Fender
1997; Eikenberry et al. 1998; Mirabel et al. 1998) was discovered, it
was suggested that the 'disappearance' of the inner disc was directly
related to its (partial) ejection (see also Feroci et al. 1999; Nandi
et al. 2001). The following sequence of events was envisaged:

\begin{enumerate}
\item{Thin disc extends close to black hole (soft state)}
\item{Inner disc is ejected, resulting in:
  \begin{itemize}
    \item{Disappearance of inner disc $\rightarrow$ transition to hard
    state}
    \item{Synchrotron event}
  \end{itemize}}
\item{Refill of disc $\rightarrow$ return to soft state}
\end{enumerate}

However, Vadawale et al. (2003) argued that it was the 'corona' which
was subsequently ejected as the disc moved in, again specifically for
the case of GRS 1915+105. Rodriguez, Corbel \& Tomsick (2003) have
also suggested, in the case of XTE J1550-564, that it is coronal, and
not disc, material which is ejected prior to radio outburst

It is clear that if the model we have outlined in this paper is
correct, the 'disc-ejection' scenario is unlikely to be, for any black
hole X-ray binary.  Specifically, it is the transition {\em towards}
the soft state (that is, the 'refill' of the inner disc) which causes
the ejection event. Therefore, if we are to consider the ejection of
mass, it is more likely the 'corona' (or whatever form the accretion
flow has in the harder states) and not the 'disc' which is ejected.

\section*{Conclusions}

We have examined the observational properties of the jets associated
with black hole X-ray binary systems. The key observations can be
summarised as:

\begin{enumerate}
\item{{\bf The radio:X-ray coupling:} we have established that the
  steady radio emission associated with the canonical LS persists
  beyond the softening of the X-ray spectrum in the 'hard' VHS/IS. At
  the end of the transtion from 'hard' to 'soft' VHS/IS, usually
  associated with a local maximum in the X-ray light curve, a
  transient radio outburst occurs. The radio emission is subsequently
  suppressed until the source X-ray spectrum hardens once more. Some
  source may repeatedly make the transition from 'hard' to 'soft'
  VHS/IS and back again, undergoing repeated episodes of steady and
  transient jet formation. }
\item{{\bf Jet velocities:} we have argued that the
  measured velocities for the transient jets, being
  relativistic with $\Gamma \ga 2$ are significantly larger than those
  of the steady jets in the LS, which probably have $\Gamma \la 1.4$.}
\item{{\bf Jet power:} we have furthermore established that our best
 estimates of the power associated with the transient jets are
 compatible with extrpolations of the functions used to estimate the
 power in the LS (albeit with a relatively large normalisation).
}
\end{enumerate}

Essentially equivalent conclusions about the radio:X-ray coupling have
been drawn by Corbel et al. (2004). Putting these observational aspects
together we have arrived at a semi-quantitative model for jet
production in black hole XRBs. We argue that for X-ray spectra harder
than some value (which may be universal or vary slightly from source
to source) a steady jet is produced. The power of this jet correlates
in a non-linear way (approximately given as $L_{\rm J} \propto L_{\rm
X}^{0.5}$) with the X-ray luminosity. As the X-ray luminosity
increases above $\sim 1$\% of the Eddington rate the X-ray spectrum
begins to soften. Physically this probably corresponds to the heating
of the inner edge of the accretion disc as it propagates inwards with
increasing accretion rate. Initially the jet production is not
affected. As the disc progresses inwards the jet velocity
increases. As it moves through the last few gravitational radii before
the ISCO, the Lorentz factor of the jet rises sharply, before the jet
is suppressed in a soft disc-dominated state. The rapid increase in
jet velocity in the final moments of its existence results in a
powerful, optically thin, internal shock in the previously existing
slower moving outflow.

The inner disc may subsequently recede, in which case a steady jet is
reformed, but with decreasing velocity and therefore no internal
shocks. If the disc once more moves inwards and reaches the 'fast jet'
zone, then once more an internal shock is formed. In fact while jets
are generally considered as 'symptoms' of the underlying accretion
flow, we consider it possible that the reverse may be true. For
example, it may be the 'growth' of the steady jet (via e.g. build up
of magnetic field near the ISCO / black hole) which results in the
hardening of the X-ray spectrum, perhaps via pressure it exerts on the
disc to push it back, or simply via Comptonisation of the inner disc
as it spreads (for further discussions see e.g. Nandi et al. 2001;
Tagger et al. 2004).

In the context of the nature and classification of black hole
'states', these states, whether 'classical' or as redefined by
McClintock \& Remillard (2004) do not have a one-to-one relation with
the radio properties of the source. It seems that as far as the jet is
concerned, it is 'on' -- albeit with a varying velocity -- if the disc
does not reach 'all the way in', which probably means as far as the
ISCO. The dividing 'jet line' (Fig 8) HID, may also correspond, at
least approximately, to a singular switch in X-ray timing properties
(Belloni et al. 2004; Homan \& Belloni 2004; see also once more
discussion in McClintock \& Remillard 2004) and may be the single
most important transition in the accretion process. Further study of
the uniqueness of the spectral and variability properties of sources
at this transition point should be undertaken to test and refine our
model.

Finally, given that Merloni, Heinz \& di Matteo (2003) and Falcke, K\"ording \&
Markoff (2004) (see also Heinz \& Sunyaev 2003; Maccarone, Gallo \&
Fender 2003) have recently demonstrated quantitatively the scaling of
radio:X-ray coupling across a range of $\ga 10^7$ in black hole mass,
it is obviously of great interest to see if the model we are working
towards for the coupling of accretion and jet formation in black hole
binaries may also be applied to AGN. In addition, detailed modelling
of the internal shock scenario is required to see if the coupling, as
outlined above, really could allow us to predict radio light curves
from X-ray, and vice versa. These two areas should be the next steps
forward.

\section*{Acknowledgements}

RPF would like to thank many people for useful discussions related to
the ideas presented here, including Catherine Brocksopp, Annalisa
Celotti, Stephane Corbel, Jeroen Homan, Peter Jonker, Marc Klein-Wolt,
Tom Maccarone, Dave Meier, Simone Migliari, Jon Miller and Felix
Mirabel. We thank the referee, Andrea Merloni, for thoughtful and
detailed comments on the paper.

\end{document}